\documentclass[conference]{IEEEtran}
\usepackage{amsmath,amssymb}
\usepackage{ascmac}
\usepackage{bm}
\usepackage[dvipdfmx,hiresbb]{graphicx}
\newtheorem{lemma}{Lemma}
\newtheorem{theorem}{Theorem}
\newtheorem{definition}{Definition}
\newtheorem{assumption}{Assumption}
\title{Continuity of the Value Function in \\ Sparse Optimal Control}
\author{\IEEEauthorblockN{Takuya Ikeda, Masaaki Nagahara}
\IEEEauthorblockA{Graduate School of Informatics, Kyoto University\\
Kyoto, 606-8501, Japan\\
Email: ikeda.t@acs.i.kyoto-u.ac.jp, nagahara@ieee.org\\
}
}
\begin{document}
\maketitle

\begin{abstract}
The purpose of this article is to show the continuity of the value function 
of the sparse optimal (or $L^0$-optimal) control problem.
The sparse optimal control is
a control whose support is minimum among all admissible controls. 
Under the normality assumption, 
it is known that a sparse optimal control is given by $L^1$ optimal control. 
Furthermore, the value function of the sparse optimal control problem is identical with 
that of the $L^1$-optimal control problem. 
From these properties, we prove the continuity of the value function 
of the sparse optimal control problem by verifying that of the  $L^1$-optimal control problem.
\end{abstract}

\section{Introduction}
\label{sec:introduction}
In this article, we consider the {\em sparse optimal control}, 
also known as the {\em maximum hands-off control}~\cite{NagQueNes13,NagQueNes14}.
A sparse control is defined as a control that has a much shorter support than the horizon length. 
A sparse optimal control is a control witch has the minimum support among all admissible controls, 
i.e., a sparse optimal control maximizes the time interval where the control value is exactly zero. 
On such a time interval, we can stop actuators.
In automobiles, for example, we can reduce $\mathrm{CO_2}$ emissions, fuel consumption, 
traffic noise and so on if we can stop actuators for long periods of time. 
Therefore the sparse optimal control has prospects for solving the environmental problems~\cite{NagQueNes14}. 

This optimal control problem is however hard to solve since the cost function is neither convex nor continuous.
To overcome this difficulty, one can adopt $L^1$ optimality as a convex approximation of the problem.
Interestingly, under a suitable assumption the solutions of the two problems are equivalent~\cite{NagQueNes13}, 
that is, a solution of the sparse optimal control problem is also one of an {\em $L^1$-optimal control problem}~\cite{Haj79}, also known as a {\em minimum fuel control problem}~\cite{AthFal}, and vice versa. 
Furthermore, the optimal values of the two problems are the same, and hence their value functions are identical.
In this article, we investigate topological properties of the value function of 
the sparse optimal control problem and prove its continuity, by using these properties.

This article is organized as follows. 
In Section \ref{sec:preliminaries}, we give mathematical preliminaries for subsequent discussion. 
In Section \ref{sec:problem}, we define the sparse optimal control problem. 
In Section \ref{sec:solutions}, we briefly review the $L^1$-optimal control, and describe the relation between the solutions of the sparse optimal control problem and those of the $L^1$-optimal control problem. 
In Section \ref{sec:value_function}, we give main theorem, that is, we prove the continuity of the value function of the sparse optimal control problem. 
Section \ref{sec:example} presents a numerical example, and we confirm the main result. 
In Section \ref{sec:conclusion}, we offer concluding remarks. 

\section{Mathematical Preliminaries}
\label{sec:preliminaries}
For $\varepsilon>0$, a set $W({\bm x},\,\varepsilon)=\{{\bm y}\in\mathbb{R}^n: \|{\bm y}-{\bm x}\|<\varepsilon\}$ 
is called the {\em $\varepsilon$-neighborhood} of ${\bm x}\in\mathbb{R}^n$, 
where $\|\cdot\|$ means the Euclidean norm.
Let $X$ be a subset of $\mathbb{R}^n$. 
A vector ${\bm x}\in X$ is called an {\em interior point} of $X$ 
if there exists $\varepsilon>0$ such that $W({\bm x},\,\varepsilon)\subset X$. 
The {\em interior} of $X$ is the set of all interior points of $X$, 
and we denote the interior of $X$ by int $X$.
A set $G\subset\mathbb{R}^n$ is said to be {\em open} 
if $G=$ int $G$. For example, int $X$ is open for every $X\subset\mathbb{R}^n$.
A vector ${\bm x}\in\mathbb{R}^n$ is called an {\em adherent point} of $X$ 
if $W({\bm x},\,\varepsilon)\cap X \neq\emptyset$ for every $\varepsilon>0$, 
and the {\em closure} of $X$ is the set of all adherent points of $X$.
A set $F\subset\mathbb{R}^n$ is said to be {\em closed} if $F=\overline{F}$, where $\overline{F}$ is the closure of $F$.
The {\em boundary} of a set $X\in\mathbb{R}^n$ is the set of all points in the closure of $X$, 
not belonging to the interior of $X$, and we denote the boundary of $X$ by $\partial X$, 
that is, $\partial X= \overline{X}-$ int $X$, 
where $X_1-X_2$ means the set of all points which belong to the set $X_1$ but not to the set $X_2$. 
In particular, if $X$ is closed, then $\partial X= X - $ int $X$, since $X=\overline{X}$.

A function $f$ defined on $\mathbb{R}^n$ is said to be {\em upper semi-continuous}
 on $\mathbb{R}^n$ if for every $\alpha\in\mathbb{R}$ the set
$\{{\bm x}\in \mathbb{R}^n: f({\bm x})<\alpha\}$
is open, and $f$ is said to be {\em lower semi-continuous} 
on $\mathbb{R}^n$ if for every $\alpha\in\mathbb{R}$ the set
$\{{\bm x}\in \mathbb{R}^n: f({\bm x})>\alpha\}$
is open. 
As a property, $f$ is continuous on $\mathbb{R}^n$ if and only if $f$ is upper and lower semi-continuous on $\mathbb{R}^n$; 
see e.g., \cite[pp. 37]{Rud}.

Let $T>0$ be fixed. 
For a continuous-time signal $u(t)$ over a time interval $[0, T]$, 
we define its {\em $L^p$ and $L^{\infty}$ norms} respectively by
\[\|u\|_{p}\triangleq\bigg\{\int_{0}^{T}|u(t)|^{p}\,dt\bigg\}^{1/p},\quad\|u\|_{\infty}\triangleq\sup_{t\in[0, T]}|u(t)|,\]
where $p\in(0,\infty)$. 
Note that if $p\in(0,\,1)$, then $\|\cdot\|_{p}$ is not a norm since it fails to satisfy the triangle inequality. 
We denote the set of all signals with $\|u\|_{p}<\infty$ by $L^p[0,\,T]$. 

We define the {\em support} of $u$, denoted by $\mathrm{supp}(u)$, as the set
\[\overline{\{t\in[0,\,T]:  u(t)\neq0\}}.\]
Then we define the {\em $L^0$ norm} of a signal $u$ as
\[\|u\|_{0}\triangleq m(\mathrm{supp}(u)),\]
where $m$ is the Lebesgue measure on $\mathbb{R}$. 
Note that the $L^0$ norm is not a norm since it fails to satisfy the positive homogeneity.
The notation $\|\cdot\|_{0}$ is justified from the fact that 
$\|u\|_{0}=\lim_{p\to 0}\|u\|_{p}^{p}$ for $u\in L^1[0,\,T]$, 
which is proved by using H\"{o}lder's inequality and Lebesgue's converge theorem~\cite{Rud}.

\section{Sparse Optimal Control Problem}
\label{sec:problem}

In this article, we will consider a linear and time-invariant control system modeled by
\begin{align}
\frac{d{\bm x}(t)}{dt}=A{\bm x}(t)+Bu(t), \tag{S}
\label{eq:system}
\end{align}
where $A$, $B$ are constant $n\times n$ and $n\times 1$ matrices respectively.
For the system \eqref{eq:system}, we call a control {\em admissible} if it steers a given initiate state ${\bm \xi}\in\mathbb{R}^n$ to the origin at fixed final time $T>0$ and is constrained in magnitude by 
\[\|u\|_{\infty}\leq1.\]
We denote by $U({\bm \xi})$ the set of all admissible controls for an initiate state ${\bm \xi}$. 
A sparse optimal control is a control that has the minimum support among all admissible controls, that is, the sparse optimal control problem for a given initiate state ${\bm \xi}$ is given as follows:
\[P_{0}:\qquad \mathrm{minimize}\,\|u\|_{0}\quad \mathrm{subject}\,\mathrm{to}\quad u\in U({\bm \xi}).\]
As described below, under a suitable assumption the solutions of this problem are those of $L^1$-optimal control problem, and vice versa~\cite{NagQueNes13}. 


\section{Solutions of Sparse Optimal Control Problem}
\label{sec:solutions}

\subsection{$L^1$-Optimal Control Problem}
The $L^1$-optimal control problem for a given initiate state ${\bm \xi}$ is described as follows:
\[P_{1}:\qquad \mathrm{minimize}\,\|u\|_{1}\quad \mathrm{subject}\,\mathrm{to}\quad u\in U({\bm \xi}).\]
This problem is also known as a {\em minimum fuel control problem}~\cite{AthFal}.
Here we briefly review the $L^1$-optimal control problem $P_1$ 
based on the discussion in \cite[Sec. 6-13]{AthFal}.

The Hamiltonian function for the $L^1$-optimal control problem is defined as
\begin{equation}
 H({\bm x},\,{\bm p},\,u)=|u|+{\bm p}^{\mathrm{T}}(A{\bm x}+Bu),
 \label{eq:Hamiltonian}
\end{equation} 
where ${\bm p}\in\mathbb{R}^n$ is the costate vector. 
Assume that $u^{\ast}$ is an $L^1$-optimal control and ${\bm x}^{\ast}$ is the resultant trajectory. According to Pontryagin's minimum principle, there exists a costate vector ${\bm p}^{\ast}$ which satisfies followings:
\begin{align*}
&H({\bm x}^{\ast},\,{\bm p}^{\ast},\,u^{\ast})\leq H({\bm x}^{\ast},\,{\bm p}^{\ast},\,u),\quad \forall u\in U({\bm \xi}),\\
&\frac{d{\bm x}^{\ast}(t)}{dt}=A{\bm x}^{\ast}(t)+Bu^{\ast}(t),\quad \frac{d{\bm p}^{\ast}(t)}{dt}=-A^{\mathrm{T}}{\bm p}^{\ast}(t),\\
&{\bm x}^{\ast}(0)={\bm \xi},\quad{\bm x}^{\ast}(T)={\bm 0}.
\end{align*}
From \eqref{eq:Hamiltonian}, the $L^1$-optimal control $u^{\ast}$ is given by 
\[u^{\ast}(t)=-\mathrm{dez}\bigl(B^{\mathrm{T}}{\bm p}^{\ast}(t)\bigr),\quad t\in[0,\,T],\]
where $\mathrm{dez}(\cdot)$ is the dead-zone function, defined by
\begin{gather*}
\mathrm{dez}(r)=\begin{cases}1,&r>1,\\0,&|r|<1,\\-1,&r<-1,\end{cases}\\
\mathrm{dez}(1)\in[0,\,1],\quad \mathrm{dez}(-1)\in[-1,\,0].
\end{gather*}

If $|B^{\mathrm{T}}{\bm p}^{\ast}(t)|$ is equal to $1$ on a time interval $[t_{1},\,t_{2}]\subset[0,\,T]$, $t_{1}<t_{2}$, then the $L^1$-optimal control $u^{\ast}(t)$ on $[t_{1},\,t_{2}]$ cannot be uniquely determined by the minimum principle. 
In this case, the interval $[t_{1},\,t_{2}]$ is called a {\em singular interval}, 
and the $L^1$-optimal control problem that has at least one singular interval is called {\em singular}.
If there exists no singular interval, the $L^1$-optimal control problem is called {\em normal}:
\begin{definition}[Normality]\label{normal}
The $L^1$-optimal control problem $P_1$ is said to be {\em normal} if the set 
\[I_0 \triangleq \{t\in[0,\,T]: |B^{\mathrm{T}}{\bm p}^{\ast}(t)|=1\}\]
is a set of measure zero, that is, $m(I_0)=0$.
\end{definition}
If the $L^1$-optimal control problem is normal, then the $L^1$-optimal control is piecewise constant and  takes vales only $\pm1$ or $0$ at almost all $t\in[0,\,T]$.

\subsection{Relation between Sparse Optimal Control and $L^1$-Optimal Control}
The following theorem describes the relation between the sparse optimal control problem $P_0$ and
the $L^1$ optimal control problem $P_1$.
\begin{theorem}
\label{theorem:relation_P0_P1}
Assume that the $L^1$-optimal control problem $P_1$ is normal and 
there exists at least one $L^1$-optimal control for a given initiate state ${\bm\xi}$. 
Let $U_{0}^{\ast}({\bm\xi})$ and $U_{1}^{\ast}({\bm\xi})$ be the sets of the optimal solutions 
of the problem $P_0$ (sparse optimal control problem) and the problem $P_1$ respectively. 
Then we have $U_{0}^{\ast}({\bm\xi})=U_{1}^{\ast}({\bm\xi})$.
Furthermore, we have $\|u_{0}\|_{0}=\|u_{1}\|_{1}$ for any $u_{0}\in U_{0}^{\ast}({\bm\xi})$
and $u_{1}\in U_{1}^{\ast}({\bm\xi})$.
\end{theorem}
\begin{IEEEproof}
By assumption, we can take any $u_{1}^{\ast}\in U_{1}^{\ast}({\bm\xi})$, and we have
\begin{align}
\begin{aligned}
\|u_{1}^{\ast}\|_{1}&=\int_{0}^{T}|u_{1}^{\ast}(t)|dt=\int_{\mathrm{supp}(u_{1}^{\ast})}|u_{1}^{\ast}(t)|dt\\
&=\int_{\mathrm{supp}(u_{1}^{\ast})}1dt=m(\mathrm{supp}(u_{1}^{\ast}))
=\|u_{1}^{\ast}\|_{0}.
\end{aligned}\label{l1_optimal}
\end{align}

Since $u_{1}^{\ast}\in U({\bm \xi})$, the set $U({\bm \xi})$ is not empty, and for any $u\in U({\bm \xi})$
we have
\begin{align}
\begin{aligned}
\|u\|_{1}&=\int_{0}^{T}|u(t)|dt=\int_{\mathrm{supp}(u)}|u(t)|dt\\
&\leq\int_{\mathrm{supp}(u)}1dt=\|u\|_{0}.
\end{aligned}\label{any}
\end{align}
From (\ref{l1_optimal}), (\ref{any}) and the optimality of $u_{1}^{\ast}$, for any $u\in U({\bm \xi})$ we have
\[\|u_{1}^{\ast}\|_{0}=\|u_{1}^{\ast}\|_{1}\leq\|u\|_{1}\leq\|u\|_{0}.\]
It follows that $u_{1}^{\ast}\in U_{0}^{\ast}({\bm\xi})$, and hence the set $U_{0}^{\ast}({\bm\xi})$ is not empty and $U_{1}^{\ast}({\bm\xi})\subset U_{0}^{\ast}({\bm\xi})$. 

On the other hands, for any $u_{0}^{\ast}\in U_{0}^{\ast}({\bm\xi})$, we have
\[\|u_{1}^{\ast}\|_{1}\leq\|u_{0}^{\ast}\|_{1}\leq\|u_{0}^{\ast}\|_{0}\leq\|u_{1}^{\ast}\|_{0}=\|u_{1}^{\ast}\|_{1}\]
by (\ref{l1_optimal}), (\ref{any}) and the optimality of $u_{0}^{\ast}$ and $u_{1}^{\ast}$. 
Therefore we have 
\begin{eqnarray}
\|u_{0}^{\ast}\|_{1}&=\|u_{1}^{\ast}\|_{1},\label{agree1}\\
\|u_{0}^{\ast}\|_{0}&=\|u_{1}^{\ast}\|_{1}.\label{agree2}
\end{eqnarray}
It follows from (\ref{agree1}) that $U_{0}^{\ast}({\bm\xi})\subset U_{1}^{\ast}({\bm\xi})$, and hence  $U_{0}^{\ast}({\bm\xi})=U_{1}^{\ast}({\bm\xi})$. Also, the last statement follows from (\ref{agree2}).\end{IEEEproof}

\section{Value Function in Sparse Optimal Control}
\label{sec:value_function}

In this section, we prove the continuity of the value function of 
the sparse optimal control problem $P_0$.

For $T\geq0$, $\alpha\geq0$, let
\[R(T)\triangleq\bigg\{\int_{0}^{T}e^{-As}Bu(s)\,ds: \|u\|_{\infty}\leq1\bigg\},\]
\[R_{\alpha}\triangleq\bigg\{\int_{0}^{T}e^{-As}Bu(s)\,ds: \|u\|_{\infty}\leq1,\,\|u\|_{1}\leq\alpha\bigg\}.\]
The set $R(T)$ is called the {\em reachable set at time $T$}.

The value function of an optimal control problem is defined as
the mapping from an initiate state to the optimal value of the cost function.
The value functions for the problems $P_0$ and $P_1$ are defined as
\begin{equation*}
V_{0}({\bm\xi})\triangleq\inf_{u\in U({\bm\xi})} \|u\|_{0},~
V_{1}({\bm\xi})\triangleq\inf_{u\in U({\bm\xi})}\|u\|_{1}.
\end{equation*}
Note that Lemma \ref{5.5} described below shows that
there exist a solution of the problem $P_1$ for any initiate state ${\bm\xi}\in R(T)$, 
and hence $V_{1}({\bm \xi})$ is well defined on $R(T)$.
Moreover,
by Theorem \ref{theorem:relation_P0_P1}, if the control problem $P_1$ is normal, 
then $V_{0}({\bm \xi})$ is also well defined on $R(T)$ and we have $V_{0}({\bm\xi})=V_{1}({\bm\xi})$ 
for any ${\bm \xi}\in R(T)$.

From these facts, we prove the continuity of $V_{0}({\bm\xi})$ on $R(T)$ by proving that of $V_{1}({\bm\xi})$. 

The next lemma is known as a sufficient condition for the $L^1$-optimal control problem to be normal~\cite{AthFal}.

\begin{lemma}\label{5.1}
If the system \eqref{eq:system} is controllable and $A$ is nonsingular, then the $L^1$-optimal control problem $P_1$ is normal.
\end{lemma}
Here we add an assumption on \eqref{eq:system} as follows:
\begin{assumption}\label{assumption}
The system \eqref{eq:system} is controllable and $A$ is nonsingular.
\end{assumption}

We then show that $V_{1}({\bm\xi})$ is continuous on $R(T)$ under Assumption \ref{assumption}.
To prove this, we need some lemmas.
\begin{lemma}\label{5.2} The followings are established:
\begin{IEEEenumerate}
\item The sets $R(T)$ and $R_{\alpha}$ are compact for $\alpha\geq0$.
\item Always, $R_{\alpha}\subset R(T)$, with equality for $\alpha\geq T$.
\item $R_0=\{{\bm 0}\}$.
\item $R_{\alpha}\subset R_{\beta}$ for $0\leq\alpha\leq\beta$.
\end{IEEEenumerate}
\end{lemma}
\begin{IEEEproof}See \cite[Lemma 2.1]{Haj79}.\qquad\end{IEEEproof}

\begin{lemma}\label{5.3}
For every $\alpha\in[0, T]$, 
\[R_\alpha=\{{\bm\xi}\in R(T):\exists\,u\in U({\bm\xi})\,\mathrm{s.t.}\,\|u\|_{1}\leq\alpha\}.\]
\end{lemma}
\begin{IEEEproof}
This follows immediately from the definition of the set $R_{\alpha}$.\end{IEEEproof}

\begin{lemma}\label{5.4}
Take any $\alpha\in[0,\,T]$.
If $u^{\ast}$ is an $L^1$-optimal control for an initiate state ${\bm\xi}\in R_{\alpha}$, then $\|u^{\ast}\|_{1}\leq\alpha.$
\end{lemma}
\begin{IEEEproof}
Fix $\alpha\in [0,\,T]$. 
Suppose that ${\bm\xi} \in R_{\alpha}$ and $u^{\ast}$ is an $L^1$-optimal control 
for the initiate state ${\bm\xi}$.
There exists a control $u\in U({\bm\xi})$ with $\|u\|_{1}\leq\alpha$ by Lemma \ref{5.3}.
Therefore we have
$\|u^{\ast}\|_{1}=V_{1}({\bm\xi})\leq\|u\|_{1}\leq\alpha$.
\end{IEEEproof}

\begin{lemma}\label{5.5}
For any initial state ${\bm \xi}\in R(T)$, 
there exists an admissible control $u$ steering the state 
from ${\bm\xi}$  to the origin at time $T$ with minimal $L^1$-cost $\|u\|_{1}$. 
Furthermore, then, ${\bm\xi}\in \partial R_{\theta}$ for $\theta=\|u\|_{1}$.
\end{lemma}
\begin{IEEEproof}See \cite[Lemma 3.1]{Haj79}.\end{IEEEproof}

\begin{lemma}\label{5.6}
For every $\alpha\in[0,\,T]$,
\[R_{\alpha}=\{{\bm\xi}\in R(T):V_{1}({\bm\xi})\leq \alpha\}.\]
\end{lemma}
\begin{IEEEproof}
Fix $\alpha\in [0,\,T]$ and take any ${\bm\xi} \in R_{\alpha}$. 
Since ${\bm\xi}\in R(T)$ by Lemma \ref{5.2}, there exists an $L^1$-optimal control $u^{\ast}$ by Lemma \ref{5.5}, and $V_{1}({\bm\xi})=\|u^{\ast}\|_{1}\leq\alpha$ by Lemma \ref{5.4}.
It follows that ${\bm \xi}\in \{{\bm\xi}\in R(T):V_{1}({\bm\xi})\leq \alpha\}$.

On the other hand, fix $\alpha\in [0,\,T]$ and take any ${\bm\xi}\in R(T)$ with $V_{1}({\bm\xi})\leq\alpha$. 
Let $V_{1}({\bm\xi})=\beta$. 
From Lemma \ref{5.5}, we have ${\bm\xi}\in\partial R_{\beta}$, and it follows from Lemma \ref{5.2} that ${\bm\xi}\in\partial R_{\beta}\subset R_{\beta}\subset R_{\alpha}$.\end{IEEEproof}

\begin{lemma}\label{5.7}
If the system \eqref{eq:system} satisfies Assumption 1, then
\[R_{\alpha}\subset \mathrm{int}\,R_{\beta}\]
whenever $0\leq\alpha<\beta\leq T$.
\end{lemma}
\begin{IEEEproof}
Let us verify only the case when $\alpha=0$. 
The other cases are proved in \cite[Lemma 4.2]{Haj79}.

Since $R_0=\{{\bm 0}\}$ by Lemma \ref{5.2}, we prove that ${\bm 0}\in$ int $R_{\beta} $ for every $\beta\in (0,\,T]$. 
Fix $\beta\in (0,\,T]$ and take an arbitrary $\gamma \in (0,\,\beta)$. 
It is already shown that $R_{\gamma}\subset$ int $R_{\beta}$. 
Since ${\bm 0}\in R_{\gamma}$, we have ${\bm 0}\in$ int $R_{\beta}$.\end{IEEEproof}

\begin{lemma}\label{5.8}
If the system \eqref{eq:system} satisfies Assumption 1, then it is necessary for every $\alpha\in[0,\,T]$ that:
\begin{IEEEenumerate}
\item \label{lem:item1} $\partial R_{\alpha}=\{{\bm\xi}\in R(T):V_{1}({\bm\xi})=\alpha\},$
\item \label{lem:item2} $\mathrm{int}R_{\alpha}=\{{\bm\xi}\in R(T):V_{1}({\bm\xi})<\alpha\}.$
\end{IEEEenumerate}
\end{lemma}
\begin{IEEEproof}
We prove the property 1; the property 2 follows immediately from the property 1 and Lemma \ref{5.6}, since $R_{\alpha}$ is closed for every $\alpha\geq0$.
If $\alpha=0$, then $\partial R_{0}=\{{\bm 0}\}$, since $R_{0}=\{{\bm 0}\}$. 
It follows from Lemma \ref{5.6} that
\[\{{\bm\xi}\in R(T):V_{1}({\bm\xi})=0\}=R_{0}=\{{\bm 0}\}=\partial R_{0}.\]
Fix $\alpha\in(0,\,T]$. 
We can take ${\bm\xi}\in\partial R_{\alpha}$, since $\partial R_{\alpha}$ is not empty. 
($\mathbb{R}^n$ and the empty set are the only subsets whose boundaries are empty, since $\mathbb{R}^n$ is connected \cite[Chapter 3]{Sin}.) 
Since ${\bm\xi}\in R_{\alpha}$, we have $V_{1}({\bm\xi})\leq\alpha$. If $V_{1}({\bm\xi})<\alpha$, then ${\bm\xi}\in\partial R_{V_{1}({\bm\xi})}\subset R_{V_{1}({\bm\xi})}\subset$ int $R_{\alpha}$, and hence a contradiction occurs. 
Therefore $V_{1}({\bm\xi})=\alpha$, and hence 
\[\partial R_{\alpha}\subset\{{\bm\xi}\in R(T):V_{1}({\bm\xi})=\alpha\}\]
and the set $\{{\bm\xi}\in R(T):V_{1}({\bm\xi})=\alpha\}$ is not empty for every $\alpha\in(0,\,T]$.
Then it follows from Lemma \ref{5.5} that 
\[\{{\bm\xi}\in R(T):V_{1}({\bm\xi})=\alpha\}\subset\partial R_{\alpha}\] 
for every $\alpha\in(0,\,T]$, and the conclusion follows.\end{IEEEproof}
Now, we prove the continuity of the value functions $V_1({\bm \xi})$
and then $V_0({\bm \xi})$.
\begin{theorem}\label{5.9}
If the system \eqref{eq:system} satisfies Assumption 1, then $V_{1}({\bm\xi})$ is continuous on $R(T)$.
\end{theorem}
\begin{IEEEproof}
Put 
\[\overline{V_1}({\bm\xi})=\begin{cases}V_{1}({\bm\xi}),&{\bm\xi}\in R(T),\\T,&{\bm\xi}\notin R(T).\end{cases}\]
It is enough to show that $\overline{V_1}({\bm\xi})$ is continuous on $\mathbb{R}^n$.

First, we show that the set 
\begin{equation}\{{\bm\xi}\in \mathbb{R}^n: \overline{V_1}({\bm\xi})<\alpha\}\label{set1}\end{equation}
is open for every $\alpha\in\mathbb{R}$
to prove $\overline{V_1}{({\bm \xi})}$ is upper semi-continuous on $\mathbb{R}^n$.
If $\alpha\leq0$ or $\alpha>T$, then the set (\ref{set1}) is empty or $\mathbb{R}^n$, respectively, and if $0<\alpha\leq T$, the set (\ref{set1}) coincides with int $R_{\alpha}$ by Lemma \ref{5.8}. 
Therefore, the set (\ref{set1}) is open for every $\alpha\in\mathbb{R}$. 
It follows that $\overline{V_1}{({\bm \xi})}$ is upper semi-continuous on $\mathbb{R}^n$.

Next, we show that the set 
\begin{equation}\{{\bm\xi}\in\mathbb{R}^n: \overline{V_1}({\bm\xi})>\alpha\}\label{set2}\end{equation}
 is open for every $\alpha\in\mathbb{R}$ to prove $\overline{V_1}({\bm\xi})$ is lower semi-continuous on $\mathbb{R}^n$.
If $\alpha<0$ or $\alpha\geq T$, then the set (\ref{set2}) coincides with $\mathbb{R}^n$ or empty, respectively, and if $0\leq\alpha<T$, from Lemma \ref{5.6}, we have
\begin{align*}
\{{\bm\xi}\in\mathbb{R}^n: \overline{V_1}({\bm\xi})>\alpha\}&=\mathbb{R}^n-\{{\bm\xi}\in R(T): V_{1}({\bm\xi})\leq\alpha\}\\
&=\mathbb{R}^n-R_{\alpha}.
\end{align*}
Therefore, the set (\ref{set2}) is open for every $\alpha\in\mathbb{R}$. It follows that $\overline{V_1}({\bm\xi})$ is lower semi-continuous on $\mathbb{R}^n$.

Hence $\overline{V_1}({\bm\xi})$ is continuous on $\mathbb{R}^n$, and the conclusion follows.\end{IEEEproof}

\begin{theorem}\label{5.10}
If the system \eqref{eq:system} satisfies Assumption 1, then $V_{0}({\bm\xi})$ is continuous on $R(T)$.
\end{theorem}
\begin{IEEEproof}
From Lemma \ref{5.5}, $V_{1}({\bm\xi})$ is well defined on $R(T)$. 
Since the $L^1$-optimal control problem is normal by Lemma \ref{5.1}, it follows from Theorem \ref{theorem:relation_P0_P1} that
$V_{0}({\bm\xi})=V_{1}({\bm\xi})$
for all ${\bm\xi}\in R(T)$, and the conclusion follows from Theorem \ref{5.9}.\end{IEEEproof}

\section{Example}
\label{sec:example}

In this section, we consider a simple example with a 1-dimensional linear control system 
\[\frac{dx(t)}{dt}=ax(t)+bu(t),\]
where $a>0$ and $b\neq0$.
Let us verify the continuity of $V_0(\xi)$ on $R(T)$. 

This system satisfies Assumption 1, and hence the sparse optimal control is given by the $L^1$-optimal control thanks to Theorem \ref{theorem:relation_P0_P1}. 
The reachable set $R(T)$ and the optimal control $u$ for an initiate state $\xi\neq0$ are computed via the bang-bang principle \cite[Theorem 12.1]{HerLas} and the minimum principle for $L^1$-optimal control \cite[Section 6.14]{AthFal} as
\[R(T)=[-x_{1},\,x_{1}],\quad u(t)=\begin{cases}-\mathrm{sgn}(b)\mathrm{sgn}(\xi),&t\in[0,\,\tau),\\0,&t\in[\tau,\,T],\end{cases}\]
where 
\[x_1=(1-e^{-aT})\frac{|b|}{a},\quad \tau=-\frac{1}{a}\log\bigl(1-\frac{|\xi|}{|b|}a\bigr),\]
and if $\xi=0$, then the optimal control takes value $0$ on $[0,\,T]$.
Then we have 
\[V_0(\xi)=\begin{cases}-\dfrac{1}{a}\log\biggl(1+\dfrac{a}{|b|}\xi\biggr),&\xi\in[-x_{1},\,0),\\0,&\xi=0,\\ -\dfrac{1}{a}\log\biggl(1-\dfrac{a}{|b|}\xi\biggr),&\xi\in(0,\,x_{1}].\end{cases}\]

\begin{figure}[!t]
\centering
   \includegraphics[width=\linewidth]{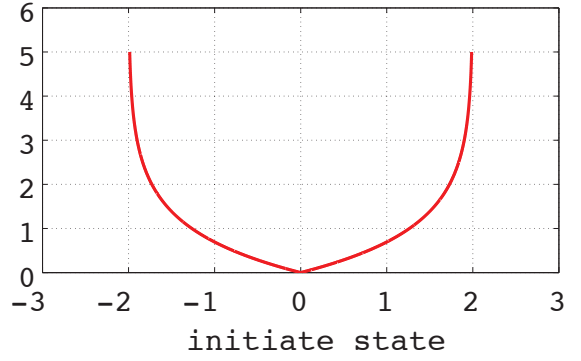}
  \caption{The value function $V_0(\xi)$}
  \label{fig:one}
\end{figure}

Fig.\ref{fig:one} shows the value function $V_{0}(\xi)$ 
for $a=1$, $b=2$, $T=5$ on $R(T)=[-2(1-e^{-5}),\,2(1-e^{-5})]$. 
Certainly, we can see that $V_{0}(\xi)$ is continuous on $R(T)$.

\section{Conclusion}
\label{sec:conclusion}
In this article, we prove the continuity of the value function of the sparse optimal control problem under the normality assumption by proving that of  the $L^1$-optimal control problem. 
The continuity of the vale function plays an important role to prove the stability when we extend it
to the model predictive control. 
An extension to the model predictive control is a future work.


\end{document}